\documentclass[twocolumn,showpacs,prd]{revtex4}
\usepackage{mathrsfs}
\usepackage{longtable,lscape}\usepackage{multirow}
\usepackage{txfonts}
\usepackage{amssymb}
\usepackage{indentfirst}
\usepackage{graphicx,,booktabs}
\usepackage{color}
\usepackage{amssymb}

\def\half{{\textstyle{1\over 2}}}

\begin{document}

\title{$Z_c(4025)$ as the hadronic molecule with hidden charm }
\author{Jun He$^{1,3}$}\email{junhe@impcas.ac.cn}
\author{Xiang Liu$^{1,2}$\footnote{Corresponding author}}\email{xiangliu@lzu.edu.cn}
\author{Zhi-Feng Sun$^{1,2}$}\email{sunzhif09@lzu.edu.cn}
\author{Shi-Lin Zhu$^4$\footnote{Corresponding author}}\email{zhusl@pku.edu.cn}
\affiliation{ $^1$Research Center for Hadron and CSR Physics,
Lanzhou University and Institute of Modern Physics of CAS, Lanzhou 730000, China\\
$^2$School of Physical Science and Technology, Lanzhou University, Lanzhou 730000,  China\\
$^3$Nuclear Theory Group, Institute of Modern Physics of CAS,
Lanzhou 730000, China\\
$^4$Department of Physics and State Key Laboratory of Nuclear
Physics and Technology, Peking University, Beijing 100871, China}

\date{\today}

\begin{abstract}

We have studied the loosely bound $D^*\bar{D}^*$ system. Our results
indicate that the recently observed charged charmonium-like
structure $Z_c(4025)$ can be an ideal $D^*\bar{D}^*$ molecular
state. We have also investigated its pionic, dipionic, and radiative
decays. We stress that both the scalar isovector molecular partner
$Z_{c0}$ and three isoscalar partners ${\tilde Z}_{c0,c1,c2}$ should
also exist if $Z_c(4025)$ is a $D^*\bar{D}^*$ molecular state
{in the framework of the one-pion-exchange model}.
$Z_{c0}$ can be searched for in the channel $e^+e^-\to Y \to
Z_{c0}(4025)\left(\pi\pi\right)_{P-wave}$ where $Y$ can be Y(4260)
or any other excited $1^{--}$ charmonium or charmonium-like states
such as Y(4360), Y(4660) etc. The isoscalar $D^*\bar{D}^*$ molecular
states ${\tilde Z}_{c0,c2}$ with $0^+(0^{++})$ and $0^+(2^{++})$ can
be searched for in the three pion decay channel $e^+e^-\to Y \to
{\tilde Z}_{c0,c2} \left(3\pi\right)^{I=0}_{P-wave}$. The isoscalar
molecular state ${\tilde Z}_{c1}$ with $0^-(1^{+-})$ can be searched
for in the channel ${\tilde Z}_{c1}\eta$. Experimental discovery of
these partner states will firmly establish the molecular picture.

\end{abstract}

 \maketitle


Since the observation of charged charmonium-like structure
$Z_c(3900)$ \cite{Ablikim:2013mio}, very recently the BESIII
Collaboration reported a new charged structure $Z_c(4025)$ in
$e^+e^-\to (D^*\bar{D}^*)^\pm\pi^\mp$ at $\sqrt{s}=4.26$ GeV
\cite{bes}. If adopting a Breit-Wigner distribution to measure
this structure, its mass and width are $(4026.3\pm2.6\pm3.7)$ MeV
and $(24.8\pm5.6\pm7.7)$ MeV \cite{bes}, respectively. In
addition, its obvious peculiarity is that $Z_c(4025)$ is near the
$D^*\bar{D}^*$ threshold. Before the observation of $Z_c(3900)$,
there are theoretical predictions of the charged charmonium-like
state near the $D^*\bar{D}^*$ threshold
\cite{Sun:2011uh,Sun:2012zzd,Chen:2011xk,chenwei}.

This new experimental observation inspires our interest in
exploring its underlying properties of $Z_c(4025)$, especially
combining our former study \cite{Sun:2011uh,Sun:2012zzd} with new
experimental information. As a charged charmonium-like structure,
$Z_c(4025)$ cannot be grouped into the conventional charmonium
family obviously. Since $Z_c(4025)$ is near the $D^*\bar{D}^*$
threshold, the exotic molecular state with hidden charm is a
possible explanation.

As a molecular candidate, $Z_c(4025)$ is composed of the $D^*$ and
$\bar{D}^*$ mesons. Since $Z_c(4025)$ appears in the
$(D^*\bar{D}^*)^\pm$ invariant mass spectrum, it should be an
isovector state. Due to the conservation of total angular momentum
and parity in the process $e^+e^-\to Y(4260)\to
Z_c(4025)^\pm\pi^\mp$, its quantum number cannot be $J^P=0^+$. The
flavor wave function of the S-wave $D^*\bar{D}^*$ molecular state
can be expressed as \cite{Sun:2011uh}
\begin{eqnarray}
\left\{\begin{array}{l}
|{Z_{c}(4025)}^+\rangle=|D^{*+}\bar{D}^{*0}\rangle\\
|{Z_{c}(4025)}^-\rangle=|D^{*-}\bar{D}^{*0}\rangle\\
|{Z_{c}(4025)}^0\rangle=\frac{1}{\sqrt{2}}\big(|D^{*+}D^{*-}\rangle-|D^{*0}\bar{D}^{*0}\rangle\big)
\end{array}\right. ,\label{e3}
\end{eqnarray}
where the total angular momentum ${J}=0,1,2$. For simplicity, we
denote these isovector molecular candidates as $Z_{cJ}$ with
${J}=0,1,2$. For the neutral state ${Z^0_{cJ}(4025)}$, its C-parity
is $C=(-)^{L+S}=(-)^J$ since $L=0$ and $S=J$. Accordingly, the
G-parity of its charged partner is $G=(-)^{J+1}$. Therefore, the
quantum numbers of the S-wave $D^*\bar{D}^*$ molecular states with
$J=0, 1, 2$ are with $I^G(J^P)=1^-(0^+),1^+(1^+),1^-(2^+)$
respectively. Since $Z_c(4025)$ was also observed in the $h_c \pi$
channel by BESIII collaboration, its G-parity is positive. In other
words, the quantum number of $Z_c(4025)$ should be
$I^G(J^P)=1^+(1^+)$ while its neutral component also carries
negative C-parity.

It's very interesting to note that the $I^G(J^{PC})=1^+(1^{+-})$
hidden-charm tetraquark states were predicted to be around 4.2 GeV
as shown in Table V of Ref. \cite{chenwei}. These tetraquark
states will fall apart into the ${\bar D}D^\ast$, ${\bar D}^\ast
D^\ast$ and other hidden-charm modes very easily. Their width is
expected to be around one or two hundred MeV. If $Z_c(4025)$ is a
tetraquark candidate, it is very challenging to explain its narrow
width and the experimental fact that such a tetraquark state does
not decay into the ${\bar D}D^\ast$ mode. Although the tetraquark
interpretation is not excluded, it seems less favorable than the
molecular picture unless one can invent a special dynamical decay
mechanism which forbids the ${\bar D}D^\ast$ mode and ensures its
narrow width. In the following we focus on the molecular
possibility.

Usually, the one-boson-exchange (OBE) model can be applied to obtain
the effective potential of $D^*$ interacting with $\bar{D}^*$. As
shown in Ref. \cite{Sun:2011uh}, the one-pion exchange is dominant
in the total effective potential of the $D^*\bar{D}^*$ system
{in the one-boson-exchange model}. Thus, in the
following we only consider the one-pion exchange contribution.

The effective Lagrangian relevant to the deduction of one pion
exchange potential include
\cite{Cheng:1992xi,Yan:1992gz,Wise:1992hn,Burdman:1992gh,Casalbuoni:1996pg,Falk:1992cx}
\begin{eqnarray}
\mathcal{L}_{HH\mathbb{P}}&=& ig\langle H^{(Q)}_b \gamma_\mu
A_{ba}^\mu\gamma_5 \bar{H}^{(Q)}_a\rangle + ig\langle
\bar{H}^{(\bar{Q})}_a \gamma_\mu
A_{ab}^\mu\gamma_5H_b^{(\bar{Q})}\rangle,\label{eq:lag}
\end{eqnarray}
where $H^{(Q)}$ and $\bar H^{(Q)}$ are expressed as
\begin{eqnarray}
H_a^{(Q)}&=&\frac{1+\rlap\slash
    v}{2}[{\mathcal{P}}^{*}_{a\mu}\gamma^\mu
    -{\mathcal{P}}_a\gamma_5],\\
    \bar{H}_a^{(Q)}&=&[{\mathcal{P}}^{*\dag}_{a\mu}\gamma^\mu
    +{\mathcal{P}}_a^{\dag}\gamma_5]\frac{1+\rlap\slash
    v}{2},
    \end{eqnarray}
where $\bar{H}=\gamma_0H^\dag\gamma_0$ and $v=(1,{\mathbf 0})$.
The multiplet field $H^{(\bar{Q})}$ with the heavy antiquark can
be defined as
\begin{eqnarray}
H_a^{(\bar{Q})}
    &=&[\widetilde{P}_a^{*\mu}\gamma_\mu-\widetilde{P}_a\gamma_5]\
    \frac{1-\rlap\slash v}{2},\\
     \bar{H}_a^{(\bar{Q})}
    &=&\frac{1-\rlap\slash
    v}{2}[\widetilde{P}_a^{*\mu}\gamma_\mu
    +\widetilde{P}_a\gamma_5].
\end{eqnarray}
Pseudoscalar ${\mathcal{P}}$ and vector ${\mathcal{P}}^{*}$ in the
multiplet field $H^{(Q)}$ are defined as ${\mathcal{P}}^{(*)T}
=(D^{(*)0},D^{(*)+})$. In the above expressions, the
$\mathcal{P}(\widetilde{\mathcal{P}})$ and
$\mathcal{P}^*(\widetilde{\mathcal{P}}^*)$ satisfy the
normalization relations $\langle
0|{\mathcal{P}}|Q\bar{q}(0^-)\rangle =\langle
0|\widetilde{\mathcal{P}}|\bar{Q}q(0^-)\rangle
=\sqrt{M_\mathcal{P}}$ and $\langle
0|{\mathcal{P}}^*_\mu|Q\bar{q}(1^-)\rangle=\langle
0|\widetilde{\mathcal{P}}^{*}_\mu|\bar{Q}q(1^-)\rangle=\epsilon_\mu\sqrt{M_{\mathcal{P}^*}}$.
The axial current is
$A^\mu=\frac{1}{2}(\xi^\dag\partial_\mu\xi-\xi \partial_\mu
\xi^\dag)=\frac{i}{f_\pi}\partial_\mu{\mathbb P}+\cdots$ with
$\xi=\exp(i\mathbb{P}/f_\pi)$ and $f_\pi=132$ MeV. Here, $\mathbb
P$ is the two by two pseudoscalar matrix.

With the Breit approximation, the one-pion exchange potential in
the momentum space can be related to the scattering amplitude by
$\mathcal{V}_\pi^{D^{*}\bar{D}^{*}}(\mathbf{q})=-{\mathcal{M}({D^{*}\bar{D}^{*}}\to
{D^{*}\bar{D}^{*}})}/{(4m^2_{D^*})}$. Thus, we obtain the
potential in the coordinate space
$\mathcal{V}_\pi^{D^{*}\bar{D}^{*}}(\mathbf{r})$ using Fourier
transformation
\begin{eqnarray}
\mathcal{V}_\pi^{D^{*}\bar{D}^{*}}(\mathbf{r})=\int\frac{d\mathbf{p}}{(2\pi)^3}\,e^{i
\mathbf{p}\cdot
\mathbf{r}}\mathcal{V}_\pi^{D^{*}\bar{D}^{*}}(\mathbf{q})\mathcal{F}^2(q^2,m_\pi^2),
\end{eqnarray}
where the monopole form factor (FF)
$\mathcal{F}(q^2,m_\pi^2)=({\Lambda^2-m_\pi^2})/({\Lambda^2-q^2})$
is introduced, which reflects the structure effect of the vertex
of the heavy mesons interacting with the light mesons. $m_\pi$ is
the exchange $\pi$ meson mass. In addition, the introduced cutoff
$\Lambda$ also plays the role of regulating the effective
potential, which is around one to several GeV.

In our calculation, we consider the S-wave and D-wave mixing
between $D^{*}$ and $\bar{D}^{*}$. Finally, the obtained one-pion
exchange potential corresponding to the different angular momentum
$\mathrm{J}$ reads (see Ref. \cite{Sun:2011uh} for more details of
the deduction)
\begin{widetext}
\begin{eqnarray}
\mathcal{V}_\pi^{D^{*}\bar{D}^{*}}(\mathbf{r})&=&
\left\{\begin{array}{lc}
\frac{g^2}{6f_\pi^2}\left[\left(\begin{array}{ccc}
2&0\\
0&-1\\
\end{array}\right)Z(\Lambda,m_\pi,r)+\left(\begin{array}{ccc}
0&\sqrt{2}\\
\sqrt{2}&2\\
\end{array}\right)T(\Lambda,m_\pi,r)\right],& \mathrm{J}=0 \\
\frac{g^2}{6f_\pi^2}\left[\left(\begin{array}{cc}
1&0\\
0&1\\
\end{array}\right)Z(\Lambda,m_\pi,r)+\left(\begin{array}{ccc}
0&-\sqrt{2}\\
-\sqrt{2}&1\\
\end{array}\right)T(\Lambda,m_\pi,r)\right],&\mathrm{J}=1
\\
\frac{g^2}{6f_\pi^2}\left[\left(\begin{array}{cccc}
-1&0&0\\
0&2&0\\
0&0&-1
\end{array}\right)Z(\Lambda,m_\pi,r)+\left(\begin{array}{cccc}
0&\sqrt{\frac{2}{5}}&-\sqrt{\frac{14}{5}}\\
\sqrt{\frac{2}{5}}&0&-\frac{2}{\sqrt{7}}\\
-\sqrt{\frac{14}{5}}&{-\frac{2}{\sqrt{7}}}&-\frac{3}{7}
\end{array}\right)T(\Lambda,m_\pi,r)\right],&\mathrm{J}=2\\
\end{array}\right. ,\label{m3}
\end{eqnarray}
\end{widetext}
where $Y(\Lambda,m_\pi,r)$, $Z(\Lambda,m_\pi,r)$ and
$T(\Lambda,m_\pi,r)$ are defined as
\begin{eqnarray}
Y(\Lambda,m_\pi,r) &=& \frac{1}{4\pi r}(e^{-m_\pi\,r}-e^{-\Lambda r})-\frac{\Lambda^2-m_\pi^2}{8\pi \Lambda }e^{-\Lambda r},\label{m1}\\
Z(\Lambda,m_\pi,r) &=& \bigtriangledown^2Y(\Lambda,m_\pi,r) = \frac{1}{r^2} \frac{\partial}{\partial r}r^2 \frac{\partial}{\partial r}Y(\Lambda,m_\pi,r),\\
T(\Lambda,m_\pi,r) &=&  r\frac{\partial}{\partial
r}\frac{1}{r}\frac{\partial}{\partial r}Y(\Lambda,m_ \pi,r).
\end{eqnarray}
Just considering the S-wave and D-wave contribution of the $D^*\bar{D}^*$ interaction,
the $D^*\bar{D}^*$ states with different ${I^G({J}^P)}$ can be expressed as
\begin{eqnarray}
|D^*\bar{D}^*[1^-(0^+)]\rangle=\left ( {\begin{array}{*{20}c}
   |D^*\bar{D}^*(^1S_0)\rangle\\
   |D^*\bar{D}^*(^5D_0)\rangle\\
\end{array}} \right ),\\
|D^*\bar{D}^*[1^+(1^+)]\rangle=\left ( {\begin{array}{*{20}c}
   |D^*\bar{D}^*(^3S_1)\rangle\\
   |D^*\bar{D}^*(^3D_1)\rangle\\
\end{array}} \right ),\\
|D^*\bar{D}^*[1^-(2^+)]\rangle=\left ( {\begin{array}{*{20}c}
   |D^*\bar{D}^*(^5S_2)\rangle\\
   |D^*\bar{D}^*(^1D_2)\rangle\\
   |D^*\bar{D}^*(^5D_2)\rangle\\
\end{array}} \right ),
\end{eqnarray}
which make the obtained potentials listed in Eq. (\ref{m3}) be of matrix form. Thus, we need to solve the coupled-channel Schr\"{o}dinger equation to find bound state solutions.

The corresponding kinetic terms are
\begin{eqnarray}
    K_{\mathrm{J}=0}&=&\mathrm{diag}\Bigg(-\frac{\triangle}{2\tilde{m}_1},~
    -\frac{\triangle_2}{2\tilde{m}_1}\Bigg),\\
    K_{\mathrm{J}=1}&=&\mathrm{diag}\Bigg(-\frac{\triangle}{2\tilde{m}_1},~
    -\frac{\triangle_2}{2\tilde{m}_1}\Bigg),\\
K_{\mathrm{J}=2}&=&\mathrm{diag}\Bigg(-\frac{\triangle}{2\tilde{m}_1},~
    -\frac{\triangle_2}{2\tilde{m}_1},~
    -\frac{\triangle_2}{2\tilde{m}_1}\Bigg),
\end{eqnarray}
where the subscripts $\mathrm{J}=0$, $\mathrm{J}=1$ and
$\mathrm{J}=2$ are introduced to distinguish the $D^*\bar{D}^*$
systems with different $\mathrm{J}$.
$\triangle=\frac{1}{r^2}\frac{\partial}{\partial
r}r^2~\frac{\partial}{\partial r}$, $\triangle_2=\triangle
-{6\over{r^2}}$. $\tilde{m}_1=m_{D^*}/2$ are the reduced mass of
the $D^*\bar{D}^*$ system, where $m_{D^{*}}$ denotes the mass of
the $D^*$ meson.

In the one-pion exchange potential, we need the value of the
coupling constant $g$ of the $D^*\bar{D}\pi$ interaction in Eq.
(\ref{m3}). In Ref. \cite{Falk:1992cx}, the coupling constant $g =
0.75$ was roughly estimated by the quark model. A different set of
coupling constants can be given in Ref. \cite{Bardeen:2003kt}.
With our notation, $g=0.6$ \cite{Bardeen:2003kt}. In fact, the
coupling constant $g$ was calculated by using many theoretical
approaches such as QCD sum rules
\cite{Belyaev:1994zk,Navarra:2001ju,Navarra:2000ji,Dai:1998vh}.
Besides these theoretical estimates of the coupling constant $g$,
$g=0.59\pm 0.07\pm0.01$ \cite{Isola:2003fh} can be extracted by
reproducing the experimental width of $D^*$ \cite{Ahmed:2001xc}.
Considering the situation of the coupling constant $g$, in this
work we discuss the dependence of binding energy on the $g$ value.
I.e., we take some typical values of the coupling constant $g$.

\begin{figure}[htb]
\begin{center}
\includegraphics[width=0.52\textwidth]{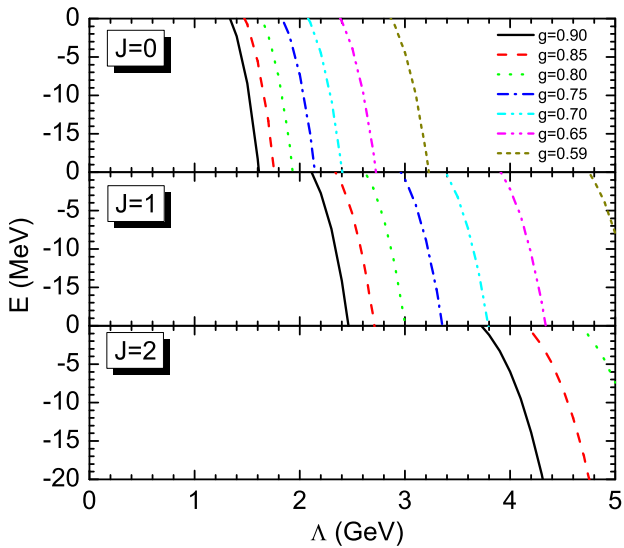}
\end{center}
\caption{The variation of the obtained binding energy of the
$D^*\bar{D}^*$ with $\mathrm{J}=0,1,2$ with the cutoff $\Lambda$.
Here, the typical values of $g$ are taken as 0.59, 0.65, 0.70,
0.75, 0.80, 0.85, 0.90. \label{Fig:E}}
\end{figure}

With the above preparation, we perform the search for the bound
state solution by solving the coupled-channel Schr\"odinger
equation. With several typical $g$ values, we present the
dependence of the binding energy $E$ on the cutoff $\Lambda$,
which is listed in Fig. \ref{Fig:E}. Since we focus on the bound
state with the low binding energy, we only give results of the
binding energy $-20\leq E\leq 0$ MeV.

If taking typical value $g=0.59$ and requiring $\Lambda<5$ GeV, we
can find bound state solutions for the isovector $D^*\bar{D}^*$ systems with
$\mathrm{J}=0,1$. If we increase the $g$ value, the corresponding
cutoff $\Lambda$ becomes smaller and is close to 1 GeV if
obtaining the same binding energy. However, for the $D^*\bar{D}^*$
system with $\mathrm{J}=2$, we cannot find a bound state solution
when taking $g=0.59$ and $\Lambda<5$ GeV. When $g=0.8$, there
exist the bound state solutions with $\Lambda\sim 5$ GeV. The
corresponding $\Lambda$ value still deviates from 1 GeV even
taking $g=0.9$. Thus, our numerical results favor the explanation
of $Z_c(4025)$ as the $D^*\bar{D}^*$ molecule with $\mathrm{J}=1$.

In addition, there also exists the $D^*\bar{D}^*$ molecular state
with $I=0$. For simplicity, we denote the isoscalar $D^*\bar{D}^*$
molecular states with $I^G(J^{PC})=0^+(0^{++})$, $0^-(1^{+-})$ and
$0^+(2^{++})$ as ${\tilde Z}_{cJ}$ where ${J}=0,1,2$. In
Ref.~\cite{Sun:2012zzd}, these isoscalar molecular states were
studied with the coupling constant $g=0.56$, and the bound state
solutions were found at $\Lambda\approx 3$~GeV, 3.5~GeV and 1.5~GeV,
respectively. If the coupling constant $g$ increases to $0.9$, the
solutions will appear at $\Lambda\approx1.5$~GeV, $1.5$~GeV and
1~GeV, respectively. Hence our calculation with the
one-pion-exchange potential supports the existence of the isoscalar
$D^*\bar{D}^*$ molecular states with $J=0,1,2$.

{The result shown in Fig. \ref{Fig:E} also indicates that the
obtained binding energy is strongly dependent on the cutoff, which
is an inherent feature of the model itself. For the molecular
states, we expect that its two constituent heavy mesons are
well-separated. Thus, the long-range interaction is very important.
However, the heavy mesons are not point-like particles. Therefore,
in the study of the molecular states with the OBE potential model
one introduces the cutoff. The role of the cutoff is to remove or
suppress the contribution from the ultraviolet region of the
exchanged momentum since the light pion sees the heavy mesons as a
whole and does not probe their inner structure.

It is well-known that the deuteron is a loosely bound molecular
state in nuclear physics. Its properties are described very well by
the meson exchange model. The binding energy of the deuteron depends
sensitively on the cutoff, the same as in our present study. Luckily
the model-dependent cutoff parameter can be extracted through
fitting to data in the deuteron case since there exist abundant
experimental data. Hopefully there may accumulate more and more
experimental data in heavy meson case in the future.

In the effective field theory framework, the scale dependence of the
loop corrections will generally be absorbed by the short-distance
counter terms in the form of low-energy constants. In the meson
exchange model, one introduces the cutoff parameter in the form
factor. The short-distance interaction is mimicked by the heavier
meson exchange such as rho and omega etc. In the present case, the
contribution from the heavier meson exchange is found to be small.
We apply the OPE model to study the newly observed $Z_c(4025)$. We
want to find whether similar bound state solutions exist for the
$D^*\bar{D}^*$ systems with the reasonable cutoff {within
the framework of the OPE model. }

In the following we discuss the decay behavior of the $D^*\bar{D}^*$
molecular states with $I^G(J^P)=1^-(0^+),1^+(1^+),1^-(2^+)$, which
is important to further test the inner structure of $Z_c(4025)$.

Under the heavy quark limit, the spin $S_H$ of the heavy quark of
the heavy-light meson is conserved as well as the total angular
momentum $J$. Thus, the spin of the light degrees of freedom,
which include light quark, gluons, quark pairs and orbit angular
momentum, can be defined as ${\vec{S}}_\ell\equiv
{\vec{J}}-{\vec{S}}_H$. The spin structure of the heavy-light
meson can be written as a form $\left[S^{P_H}_H\otimes
    S^{P_\ell}_\ell\right]_{J^P}$, where $P_H$ and $P_\ell$ are
the parities of heavy quark and the remaining light degrees of
freedom in heavy-light meson, which satisfy $P_HP_\ell=P$. For
example, the corresponding spin structure for the $D^*$ meson is
$\left[\frac{1}{2}^+_H\otimes\frac{1}{2}^-_\ell\right]_{1^-}$~\cite{Ohkoda:2012rj}.
Thus, the S-wave $D^*\bar{D}^*$ molecular systems with $J=0,1,2$
can be expressed by the spin re-coupling formula with $9-j$
symbols \cite{Ohkoda:2012rj,Liu:2013rxa}, i.e.,
\begin{eqnarray}
     &&D^*\bar{D}^*[J^P]\Rightarrow\left[\left[\half_H^+\otimes\half^-_\ell\right]_{1^-}~\left[\half_H^-\otimes\half^+_\ell\right]_{1^-}\right]_{J^P}\nonumber\\
    &&=\sum_{S_{\hat{H}},S_{\hat{\ell}}}3\sqrt{2S_{\hat H}+1}\sqrt{2S_{\hat \ell}+1}
    \left\{\begin{array}{ccc}
        1/2&1/2&1\\
        1/2&1/2&1\\
        S_{\hat{H}}&S_{\hat{\ell}}&J
    \end{array}\right\}\left[S^-_{\hat{H}}\otimes S_{\hat\ell}^-\right]_{J^P}\nonumber\\
&&=\left\{\begin{array}{ll}
{\sqrt{3}\over 2}\left[{0^-_{\hat{H}}}\otimes {0_{\hat\ell}^-}\right]_{0^+}-{1\over 2}\left[{1^-_{\hat{H}}}\otimes {1^-_{\hat\ell}}\right]_{0^+}, & J=0\\
{\sqrt{2}\over 2}\left[{0^-_{\hat{H}}}\otimes {1_{\hat\ell}^-}\right]_{1^+}+{\sqrt{2}\over2}\left[ {1^-_{\hat{H}}}\otimes {0^-_{\hat\ell}}\right]_{1^+},& J=1\\
\left[{1^-_{\hat{H}}}\otimes {1^-_{\hat\ell}} \right]_{2^+}, &J=2
\end{array}\right.   ,\label{recoupling}
\end{eqnarray}
where ${S_{\hat{H}}}=\vec{\frac{1}{2}}_H+\vec{\frac{1}{2}}_H$
denotes the angular momentum composed of the spins of $c$ and
$\bar{c}$ quarks. Thus, $S_{\hat{H}}=0,1$. The subscript in
$S_{\hat{H}}^-$ denote the parity constructed by the parities of
$c$ and $\bar{c}$ quarks directly. $S_\ell$ denotes the angular
momentum of the remaining light degrees of freedom in the
$D^*\bar{D}^*$ molecular state, which is similar to the definition
of $S_\ell$. Since the parity of the S-wave $D^*\bar{D}^*$
molecular state is constrained to be $+$, thus the parity of light
degrees of freedom in the $D^*\bar{D}^*$ molecular state must be
$-$, where we add the subscript $-$ to $S_{\hat{\ell}}$ in Eq.
(\ref{recoupling}).

In Table~\ref{Tab: decay}, the spin structures of the molecular
states and the charmoniums are presented explicitly. The angular
momentum and parity $J^P$ of the emitted pion or photon can be
obtained from the spin structures of the initial and final states
as shown in Table ~\ref{Tab: decay}. Here, the second column is
the corresponding information of $\left[S^-_{\hat{H}}\otimes
S_{\hat\ell}^-\right]_{J^P}$ of $Z_c(4025)$ with different $J^P$
quantum numbers as shown in Eq. (\ref{recoupling}). The final
states of the pionic and radiative decays of $Z_c(4025)$ are given
in the first row. In addition, we list the spin structure of final
states of the corresponding decays in the second row, where we
also use the notation $\left[S^-_{\hat{H}}\otimes
S_{\hat\ell}^-\right]_{J^P}$ for charmonium, which is similar to
that for the $D^*\bar{D}^*$ molecular state. We need to specify
that the definitions of $S_{\hat\ell}$ for charmonium and the
$D^*\bar{D}^*$ molecular state are slightly different.
$S_{\hat\ell}$ for the charmonium denotes the angular momentum of
the remaining degrees of freedom when excluding heavy quarks in
charmonium, while $S_{\hat\ell}$ for the $D^*\bar{D}^*$ molecular
state contains the degrees of freedom of the light valence quarks.
For the pion and photon, the former definition
$\left[S^-_{\hat{H}}\otimes S_{\hat\ell}^-\right]_{J^P}$ is
abbreviated as $J^P_\ell$. In Table \ref{Tab: decay}, we use $--$
to mark it if the corresponding decay channel is forbidden. In
addition, for those allowed decay channels, we further give the
information of the $J^P$ quantum numbers carried by the pion or
photon, which is different from the intrinsic spin-parity of pion
or photon. The indices $S$ and $P$ denote the allowed orbital
angular momentum between charmonium and light meson for the
corresponding decay.

\begin{table*}[htbp!]
\renewcommand\tabcolsep{0.12cm}
\renewcommand{\arraystretch}{1.8}

\caption{The pionic and radiative decays of the $D^*\bar{D}^*$
molecular states with different ${I^G(J^P)}$ in the heavy quark limit.
    \label{Tab: decay}}
\begin{tabular}{cr|cc|cc|cc|cc|cc|cc}\toprule[1pt]
    & & $h_c(1P)$& $\pi$  & $J/\psi$ & $\pi$ & $\eta_c$ & $\gamma$
    &$\chi_{c0}$ & $\gamma$&$\chi_{c1}$ & $\gamma$&$\chi_{c2}$ & $\gamma$

\\
& &$\left[0^-_{\hat{H}}\otimes1^-_\ell\right]_{1^+}$&$0^-_\ell$
&$\left[1^-_{\hat{H}}\otimes
    0^+_\ell\right]_{1^-}$ & $0^-_\ell$ & $\left[0^-_{\hat{H}}\otimes
        0^+_\ell\right]_{0^-}$
    &$1^-_\ell$&$\left[1^-_{\hat{H}}\otimes 1^-_\ell\right]_{0^+}$ &    $1^-_\ell$ &$\left[1^-_{\hat{H}}\otimes 1^-_\ell\right]_{1^+}$&
        $1^-_\ell$ &$\left[1^-_{\hat{H}}\otimes 1^-_\ell\right]_{2^+}$ &    $1^-_\ell$
\\\midrule[1pt]
    \multirow{2}*{$D^*\bar{D}^*[1^-(0^+)]$}
    &$\frac{\sqrt{3}}{2}\left[0^-_{\hat{H}}\otimes0^-_{\hat{\ell}}\right]_{0^+}$
    &\multicolumn{2}{c|}{$--$}
    &\multicolumn{2}{c|}{$--$}&\multicolumn{2}{c|}{$--$}&\multicolumn{2}{c|}{$--$}&\multicolumn{2}{c|}{$--$}&\multicolumn{2}{c}{$--$}\\
    &$-\frac{1}{2}\left[1^-_{\hat{H}}\otimes1^-_{\hat{\ell}}\right]_{0^+}$&\multicolumn{2}{c|}{$--$}&\multicolumn{2}{c|}{$--$}&\multicolumn{2}{c|}{$--$}&\multicolumn{2}{c|}{$--$}
&\multicolumn{2}{c|}{$--$}&\multicolumn{2}{c}{$--$}\\
\midrule[1pt]
    \multirow{2}*{$D^*\bar{D}^*[1^+(1^+)]$}
    &$\frac{1}{\sqrt{2}}\left[0^-_{\hat{H}}\otimes1^-_{\hat{\ell}}\right]_{1^+}$&\multicolumn{2}{c|}{$1^+\
    (P)$}
&\multicolumn{2}{c|}{$--$}&\multicolumn{2}{c|}{$1^-$}&\multicolumn{2}{c|}{$--$}&\multicolumn{2}{c|}{$--$}&\multicolumn{2}{c}{$--$}\\
&$\frac{1}{\sqrt{2}}\left[1^-_{\hat{H}}\otimes0^-_{\hat{\ell}}\right]_{1^+}$&\multicolumn{2}{c|}{$--$}
&\multicolumn{2}{c|}{$0^-\
(S)$}&\multicolumn{2}{c|}{$--$}&\multicolumn{2}{c|}{$1^+$}&\multicolumn{2}{c|}{$1^+$}&\multicolumn{2}{c}{$1^+$}\\
\midrule[1pt]
    \multirow{1}*{$D^*\bar{D}^*[1^-(2^+)]$}
&$\left[1^-_{\hat{H}}\otimes1^-_{\hat{\ell}}\right]_{2^+}$&\multicolumn{2}{c|}{$--$}
&\multicolumn{2}{c|}{$--$}&\multicolumn{2}{c|}{$--$}&\multicolumn{2}{c|}{$--$}&\multicolumn{2}{c|}{$--$}&\multicolumn{2}{c}{$--$}\\
\midrule[1pt] \multirow{2}*{$Z_{c}(3900)[1^+(1^+)]$}
&$-\frac{1}{\sqrt{2}}\left[0^-_{\hat{H}}\otimes1^-_{\hat{\ell}}\right]_{1^+}$
&\multicolumn{2}{c|}{$1^+\ (P)$}
&\multicolumn{2}{c|}{$--$}&\multicolumn{2}{c|}{$1^-$}&\multicolumn{2}{c|}{$--$}&\multicolumn{2}{c|}{$--$}&\multicolumn{2}{c}{$--$}\\
&$\frac{1}{\sqrt{2}}\left[1^-_{\hat{H}}\otimes0^-_{\hat\ell}\right]_{1^+}$
&\multicolumn{2}{c|}{$--$} &\multicolumn{2}{c|}{$0^-\
(S)$}&\multicolumn{2}{c|}{$--$}&\multicolumn{2}{c|}{$1^+$}&\multicolumn{2}{c|}{$1^+$}&\multicolumn{2}{c}{$1^+$}\\
\midrule[1pt]
\end{tabular}
\end{table*}

Besides the angular momentum and parity conservations, the most
important rule is that the decay with heavy quark spin flip should
be suppressed when the heavy quark mass approaches infinity, i.e.,
$1^-_H\nrightarrow 0^-_H$ is forbidden. The $h_c(1P)\pi$ and
$J/\psi\pi$ decay modes are forbidden for the $D^*\bar{D}^*$
molecular states with $J^P=0^+$ and $2^+$ due to the G-parity
conservation. Besides these qualitative conclusion of the
$D^*\bar{D}^*$ molecular state decay into $h_c(1P)\pi$ and $J/\psi
\pi$, in the following we give a quantitative discussion of these
decays associated with $Z_c(3900)$, where $Z_c(3900)$ is another
charged charmonium-like structure observed by BESIII in $e^+e^-\to
J/\psi\pi^+\pi^-$ at $\sqrt{s}=4.26$ GeV \cite{Ablikim:2013mio}.
$Z_c(3900)$ can be explained as the $D\bar{D}^*$ molecular state
\cite{Wang:2013cya,Guo:2013sya,Voloshin:2013dpa,Zhang:2013aoa},
where  $Z_c(3900)\to h_c(1P)\pi$ and $Z_c(3900)\to J/\psi\pi$ is
also allowed (see Table \ref{Tab: decay} for its decay
information).

With Eq. (\ref{recoupling}) and Table \ref{Tab: decay}, the decay
width ratio of the $h_c(1P)\pi$ and $J/\psi\pi$ modes of $Z_c(4025)$
and $Z_c(3900)$ under the heavy quark limit is
\begin{eqnarray}
\Gamma\left[Z_c(4025)\right]:\Gamma\left[Z_c(3900)\right]=1, \label{ha}
\end{eqnarray}
where we have ignored the phase space difference for simplicity.
{The above ratio is only roughly estimated without
involving any dynamics. If considering the concrete dynamics, this
ratio may change. Thus, a further study on this ratio is needed with
some special models, which will be another interesting research
topic. }

Since BESIII announced $Z_c(3900)\to J/\psi\pi$
\cite{Ablikim:2013mio} and has not reported any enhancement
structure near the $D^*\bar{D}^*$ threshold up to now, one may
wonder why the $Z_c(4025)$ signal is absent in the $J/\psi$
invariant mass spectrum of $e^+e^-\to J/\psi\pi^+\pi^-$ process at
$\sqrt{s}=4.26$ GeV if $Z_c(4025)$ is a $D^*\bar{D}^*$ molecular
state with $J^P=1^+$. One possibility is that the phase space of
$Y(4260)\to Z_c(4025) \pi$ is much smaller than that $Y(4260)\to
Z_c(3900)\pi$. Thus, the signal of $Z_c(4025)$ with $J^P=1^+ $
might be buried by that of $Z_c(3900)$ in the $e^+e^-\to
J/\psi\pi^+\pi^-$ process.

Additionally, we also provide the radiative decay information of
$D^*\bar{D}^*$ molecular states. The decay of the $D^*\bar{D}^*$
molecular state with $J^P=0^+$ into $\chi_{c0}$ should be
suppressed because it is a $0\to0$ process.  However, for the
$D^*\bar{D}^*$ molecular state with $J^P=1^+$, its radiative
decays into $\chi_{cJ}$ ($J=0,1,2$) are the typical $M1$
transition. Our calculation shows
\begin{eqnarray}
\Gamma(\chi_{c0}\gamma):\Gamma(\chi_{c1}\gamma):\Gamma(\chi_{c2}\gamma)
=1:3:5.
\end{eqnarray}
After considering the phase space factors proportional to the
cubic of the photon energy, the above ratio becomes
$$\Gamma(\chi_{c0}\gamma):\Gamma(\chi_{c1}\gamma):\Gamma(\chi_{c2}\gamma)
=1:1.9:2.4.$$ Experimental study of these radiative decays of
$Z_c(4025)$ is an interesting topic, which will be helpful to test
the quantum number assignment of $Z_c(4025)$.

We also present the hidden-charm dipion decays of the
$D^*\bar{D}^*$ molecular state as shown in Table~\ref{Tab:
2pidecay}. The $h_c(1P)[\pi\pi]_P$ and $J/\psi[\pi\pi]_P$ decay
modes are allowed for the $D^*\bar{D}^*$ molecular states with
$J^P=0^+$ and $2^+$, where the subscript $P$ denotes the relative
angular momentum between the two pions. The ratio of the decays of
the $D^*\bar{D}^*$ molecular state with $J^P=1^+$ into
$\chi_{cJ}[\pi\pi]_P$  ($J=0,1,2$) should be similar to that of
the corresponding radiative decays due to the similar quantum
number and phase space factor.

\begin{table*}[htbp!]
\renewcommand\tabcolsep{0.24cm}
\renewcommand{\arraystretch}{1.8}
\caption{{The hidden-charm dipion decays of the $D^*\bar{D}^*$
molecular states with different ${I^G(J^P)}$.  \label{Tab: 2pidecay}}}
\begin{tabular}{cr|cc|cc|cc|cc|cc|cc}\toprule[1pt]
    & & $h_c(1P)$& $[\pi\pi]_P$  & $J/\psi$ & $[\pi\pi]_P$ & $\eta_c$ & $[\pi\pi]_P$
    &$\chi_{c0}$ &$[\pi\pi]_P$&$\chi_{c1}$ & $[\pi\pi]_P$&$\chi_{c2}$ & $[\pi\pi]_P$
\\\midrule[1pt]
    \multirow{2}*{$D^*\bar{D}^*[1^-(0^+)]$}
    &$\frac{\sqrt{3}}{2}\left[0^-_{\hat{H}}\otimes0^-_{\hat{\ell}}\right]_{0^+}$
    &\multicolumn{2}{c|}{$1^+$}
    &\multicolumn{2}{c|}{$--$}&\multicolumn{2}{c|}{$--$}&\multicolumn{2}{c|}{$--$}&\multicolumn{2}{c|}{$--$}&\multicolumn{2}{c}{$--$}\\
    &$-\frac{1}{2}\left[1^-_{\hat{H}}\otimes1^-_{\hat{\ell}}\right]_{0^+}$&\multicolumn{2}{c|}{$--$}&\multicolumn{2}{c|}{$1^-$}&\multicolumn{2}{c|}{$--$}&\multicolumn{2}{c|}{$--$}
&\multicolumn{2}{c|}{$--$}&\multicolumn{2}{c}{$--$}\\
\midrule[1pt]
    \multirow{2}*{$D^*\bar{D}^*[1^+(1^+)]$}
    &$\frac{1}{\sqrt{2}}\left[0^-_{\hat{H}}\otimes1^-_{\hat{\ell}}\right]_{1^+}$&\multicolumn{2}{c|}{$--$}
&\multicolumn{2}{c|}{$--$}&\multicolumn{2}{c|}{$1^-$}&\multicolumn{2}{c|}{$--$}&\multicolumn{2}{c|}{$--$}&\multicolumn{2}{c}{$--$}\\
&$\frac{1}{\sqrt{2}}\left[1^-_{\hat{H}}\otimes0^-_{\hat{\ell}}\right]_{1^+}$&\multicolumn{2}{c|}{$--$}
&\multicolumn{2}{c|}{$--$}&\multicolumn{2}{c|}{$--$}&\multicolumn{2}{c|}{$1^+$}&\multicolumn{2}{c|}{$1^+$}&\multicolumn{2}{c}{$1^+$}\\
\midrule[1pt]
    \multirow{1}*{$D^*\bar{D}^*[1^-(2^+)]$}
&$\left[1^-_{\hat{H}}\otimes1^-_{\hat{\ell}}\right]_{2^+}$&\multicolumn{2}{c|}{$--$}
&\multicolumn{2}{c|}{$1^-$}&\multicolumn{2}{c|}{$--$}&\multicolumn{2}{c|}{$--$}&\multicolumn{2}{c|}{$--$}&\multicolumn{2}{c}{$--$}\\
\midrule[1pt] \multirow{2}*{$Z_{c}(3900)[1^+(1^+)]$}
&$-\frac{1}{\sqrt{2}}\left[0^-_{\hat{H}}\otimes1^-_{\hat{\ell}}\right]_{1^+}$
&\multicolumn{2}{c|}{$--$}
&\multicolumn{2}{c|}{$--$}&\multicolumn{2}{c|}{$1^-$}&\multicolumn{2}{c|}{$--$}&\multicolumn{2}{c|}{$--$}&\multicolumn{2}{c}{$--$}\\
&$\frac{1}{\sqrt{2}}\left[1^-_{\hat{H}}\otimes0^-_{\hat\ell}\right]_{1^+}$
&\multicolumn{2}{c|}{$--$} &\multicolumn{2}{c|}{$--$}&\multicolumn{2}{c|}{$--$}&\multicolumn{2}{c|}{$1^+$}&\multicolumn{2}{c|}{$1^+$}&\multicolumn{2}{c}{$1^+$}\\
\midrule[1pt]
\end{tabular}
\end{table*}

Similarly for the isoscalar $D^*\bar{D}^*$ molecular states with
$0^+(0^{++})$, $0^-(1^{+-})$ and $0^+(2^{++})$, the possible hidden-charm decay channels are
\begin{eqnarray}
{\tilde Z}_{c0}&\to& J/\psi\omega,~J/\psi\gamma,~\eta_c\eta,\nonumber\\
{\tilde Z}_{c1}&\to& \eta_c\omega,~ \eta_c\gamma,~J/\psi\eta,~
\chi_{cJ}\gamma\nonumber\\
{\tilde Z}_{c2}&\to& J/\psi\omega,~J/\psi\gamma.\nonumber
\end{eqnarray}
The decay of ${\tilde Z}_{c2} \to  \eta_c\eta$ is suppressed due to
the heavy quark spin flip.

In summary, stimulated by the observed $Z_c(4025)$ by BESIII
\cite{bes}, we have investigated whether $Z_c(4025)$ can be
explained as the $D^*\bar{D}^*$ molecular state by performing the
calculation of the mass spectrum and the analysis of its pionic and
radiative decays. The effective potential of the $D^*\bar{D}^*$
interaction can be obtained by the one-pion exchange model. By
solving the Schr\"{o}dinger equation, we find the bound state solutions
with a reasonable $\Lambda$ value for the $D^*\bar{D}^*$ molecular
states with $J^P=0^+,1^+$, which indicates that $Z_c(4025)$ can be the
ideal candidate of the $D^*\bar{D}^*$ molecular states with
$J^P=1^+$. Under the heavy quark limit, we have also studied the
decay behavior of these $D^*\bar{D}^*$ molecular states.
Experimental measurement of these ratios will probe the internal
structure of $Z_c(4025)$.

{In addition, both the scalar isovector molecular partner $Z_{c0}$ and three isoscalar partners ${\tilde Z}_{c0,c1,c2}$ should also exist if $Z_c(4025)$ is a $D^*\bar{D}^*$ molecular state within the one-pion exchange formalism. Their possible existence is strongly linked to the one-pion exchange (OPE) potential assumed in the calculation. The discovery of these additional states would give strong support of the OPE model and the molecular interpretation for $Z_c(4025)$.}
Experimental discovery of these partner states will
establish the molecular picture.

The molecular state $Z_{c0}(4025)$ can be searched for in the
channel $e^+e^-\to Y \to Z_{c0}(4025)\left(\pi\pi\right)_{P-wave}$
where $Y$ can be Y(4260) or any other excited $1^{--}$ charmonium or
charmonium-like states such as Y(4360), Y(4660) etc. Here the dipion
acts as the quantum number filter. The isoscalar $D^*\bar{D}^*$
molecular state with $0^+(0^{++})$ and $0^+(2^{++})$ can be searched
for in the three pion decay channel $e^+e^-\to Y(4260, 4360,
4660)\mbox{etc} \to {\tilde Z}_{c0,c2}
\left(3\pi\right)^{I=0}_{P-wave}$. The isoscalar $D^*\bar{D}^*$
molecular state with $0^-(1^{+-})$ can be searched for in the
channel ${\tilde Z}_{c1}\eta$ through the same process.

More and more experimental progress on charged charmonium-like
states provide us with a good platform to study exotic hadrons, which is a
research field full of challenges and opportunities. More
theoretical and experimental efforts are called for in order to
reveal the underlying mechanism behind these novel phenomena.

\section*{Acknowledgement}

This project is supported by the National Natural Science
Foundation of China (Grants No. 11275235, No. 11075004, No.
11021092, No. 11035006, No. 11047606, No. 10805048), and the
Ministry of Science and Technology of China (No. 2009CB825200),
and the Ministry of Education of China (FANEDD under Grants No.
200924, DPFIHE under Grants No. 20090211120029, NCET under Grants
No. NCET-10-0442, the Fundamental Research Funds for the Central
Universities under Grants No. lzujbky-2010-69), the Knowledge
Innovation Project of the Chinese Academy of Sciences (Grant No.
KJCX2-EW-N01).

\end{document}